\documentclass[12pt]{article}
\usepackage{xcolor, soul}
\sethlcolor{red}
\usepackage{graphicx}
\usepackage{hyperref}
\usepackage{float}
\usepackage{amsmath}
\usepackage[title]{appendix}
\usepackage{cite}
\usepackage{authblk}
\usepackage{caption}
\usepackage{bbold}
\usepackage{subfigure}

\begin{document}
\title{A randomly generated Majorana neutrino mass matrix using Adaptive Monte Carlo method}
\author{Y Monitar Singh$^{1}$\footnote{monitar.phd.phy@manipuruniv.ac.in}, Mayengbam Kishan Singh$^{1,2}$\footnote{kishan@manipuruniv.ac.in} and N Nimai Singh$^{1,2}$\footnote{nimai03@yahoo.com}\\
 $^{1}$\small Department of Physics, Manipur University, Imphal-795003, India\\
 $^{2}$\small Research Institute of Science and Technology, Imphal-795003, India}
\date{}
\maketitle
\begin{abstract}
 A randomly generated complex symmetric matrix using Adaptive Monte Carlo method, is taken as a general form of Majorana neutrino mass matrix, which is diagonalized by the use of eigenvectors.  We extract all the neutrino oscillation parameters i.e. two mass-squared differences ($\Delta m_{21}^2$ and $\Delta m_{32}^2$ ), three mixing angles ($\theta_{12}$, $\theta_{13}$, $\theta_{23}$) and three phases i.e. one Dirac CP violating phase ($\delta_{CP}$) and two Majorana phases ($\alpha_1$ and $\alpha_2$). The charge-parity (CP) violating phases are extracted from the mixing matrix constructed with the eigenvectors of the Hermitian matrix formed by the complex symmetric mass matrix. All the latest neutrino oscillation parameters within 3$\sigma$ bound are only allowed in normal hierarchy (NH) consistent with the latest Planck cosmological upper bound, $\sum\vert m_i\vert<0.12$ eV. Three cases of one-zero texture i.e. $m_{11}=0$, $m_{12}=0$, $m_{13}=0$ and two cases of two-zero texture i.e. $m_{11},m_{12}=0$ and $m_{11},m_{13}=0$ are allowed in normal hierarchy whereas none of zero texture is allowed in inverted hierarchy.  We also study effective neutrino masses $m_{\beta}$ in tritium beta decay and $m_{\beta\beta}$ in neutrinoless double beta decay and calculate numerically the allowed range of mass elements of Majorana mass matrix.  
\end{abstract}

\section{Introduction} 

The present neutrino oscillation data confirms that neutrinos have very tiny but non-zero masses. These tiny masses of neutrinos are elegantly explained by the celebrated seesaw mechanism \cite{akhmedov2000seesaw,dienes1999light}. Attempts have been made \cite{de2016neutrino,cai2017trees,king2014neutrino,gonzalez2003neutrino} to construct possible neutrino mass matrices, including zero-texture models. However, the ambiguity in the choice of the correct form of this texture, is not yet resolved. With the hope to discriminate among these theoretically predicted possible textures, we adopt a different approach that may give some hints on the correct structure of the Majorana neutrino mass matrix consistent with the presently available neutrino oscillation data. One such approach is the Adaptive Monte Carlo method for randomly generating the complex Majorana neutrino mass matrix \cite{dziewit2006majorana,kiers2002ubiquitous}.

A Majorana mass matrix, $m_\nu$ can be constructed by using a set of six real numbers as well as a set of six complex numbers. A set of real numbers can form only a symmetric Majorana mass matrix for charge-parity (CP) conserving case with Dirac phase, $\delta_{CP}=0$. This Majorana neutrino mass matrix is diagonalized the use of the standard mixing matrix, $U$ as follows, 

\begin{equation}
M_{diag}= diag(m_1, m_2, m_3)=U^T m_\nu U
\end{equation} 
 
For CP non-conserving case with $\delta_{CP}\neq 0$, a 3$\times$3 complex symmetric matrix is taken as a general form of Majorana mass matrix, $m_\nu$. Using this complex symmetric mass matrix, one can construct a Hermitian matrix $h$ as $h=m_\nu m_\nu^{\dag}$, which is digonalized as \cite{adhikary2013masses}

\begin{equation}
U^{\dag}hU=diag(m_1^2, m_2^2, m_3^2)=D
\end{equation}

where the $U$ is mixing matrix constructed with the eigenvectors of $h$ and $m_i^2$ (i=1,2,3) are the mass-squared eigenvalues.

The general form of Majorana neutrino mass matrix has the following structure:

\begin{equation}
m_{\nu}=\left(\begin{array}{ccc}
a&b&c\\
b&d&e\\
c&e&f
\end{array}\right)=\left(\begin{array}{ccc}
m_{11}e^{i\phi_{11}}&m_{12}e^{i\phi_{12}}&m_{13}e^{i\phi_{13}}\\
m_{12}e^{i\phi_{12}}&m_{22}e^{i\phi_{22}}&m_{23}e^{i\phi_{23}}\\
m_{13}e^{i\phi_{13}}&m_{23}e^{i\phi_{23}}&m_{33}e^{i\phi_{33}}
\end{array}\right)
\end{equation}

One of the major reasons for considering complex symmetric matrix is to extract the charge-parity (CP) violating one Dirac phase and two Majorana phases. A real symmetric matrix gives only real eigenvectors whereas a complex symmetric matrix gives complex eigenvectors. The elements of mixing matrix, $U$ constructed by eigenvectors, should also be complex numbers which lead to CP-violating phases. The general form of mixing matrix, $U$ takes the following form \cite{gupta2015renormalization}

\begin{align}
U=\left(\begin{array}{ccc}
c_{13}c_{12}&c_{13}s_{12}&s_{13}e^{-i\delta_{CP}}\\
-c_{23}s_{12}-c_{12}s_{13}s_{23}e^{-i\delta_{CP}}&c_{12}c_{23}-s_{12}s_{13}s_{23}e^{-i\delta_{CP}}&c_{13}s_{23}\\
s_{12}s_{23}-c_{12}s_{13}c_{23}e^{-i\delta_{CP}}&-c_{12}s_{23}-c_{23}s_{13}s_{12}e^{-i\delta_{CP}}&c_{13}c_{23}
\end{array}\right).P
\end{align}

where $s_{ij}=\sin\theta_{ij}$, $c_{ij}=\cos\theta_{ij}$; $i,j=1,2,3$, $i\neq j$ and $\delta_{_{CP}}$=Dirac CP violating phase. The matrix $P=diag(e^{-i\alpha_1},e^{-i\alpha_2}, 1)$ has two Majorana CP phases $\alpha_1$ and $\alpha_2$. From the complex mixing matrix $U$ and $D$, we can extract all neutrino oscillation data including the Dirac and Majorana CP-violating phases by using the following relations \cite{jarlskog1985commutator}:
$\Delta m_{21}^2=m_2^2-m_1^2$;
$\Delta m_{32}^2=m_3^2-m_2^2$;
$\tan\theta_{12}=\frac{|U_{e2}|}{|U_{e1}|}$;
$\tan\theta_{23}=\frac{|U_{\mu3}|}{|U_{\tau3}|}$;
$\sin\theta_{13}=|U_{e3}|$. The tangent of mixing angles are also converted into sine of mixing angles as $\sin\theta_{ij}=\frac{\tan\theta_{ij}}{\sqrt{1+\tan^2\theta_{ij}}}$ in the numerical analysis for convenience.

For one Dirac CP-phase \cite{jarlskog1985commutator},
\begin{equation}
J_{CP}=Im[U_{e1}U_{\mu2}U_{e2}^{*}U_{\mu}^{*}]=s_{12}c_{12}s_{23}c_{23}c_{13}s_{13}\sin\delta_{_{CP}}
\end{equation}

For two Majorana phases \cite{jarlskog1985commutator},
\begin{equation}
I_1=Im[U_{e1}^{*}U_{e2}]=c_{12}c_{13}^2s_{12}\sin(\frac{\alpha_1}{2})
\end{equation}

\begin{equation}
I_2=Im[U_{e1}^{*}U_{e3}]=c_{12}c_{13}s_{13}\sin(\frac{\alpha_2}{2}-\delta_{CP})
\end{equation}
where $J_{_{CP}}$ is the Jarlskog rephasing invariant quantity that measures the magnitude of Dirac CP violation, and  $I_1$ and $I_2$ are other two invariant quantities that measure the magnitude of Majorana CP violation.

The present work is an attempt to construct a possible Majorana neutrino mass structure that can give all neutrino oscillation data within 3$\sigma$ bound \cite {gonzalez2021nufit,esteban2018updated,esteban2020fate}. There are generally two popular approaches to restrict the form of neutrino mass matrices namely, ``top-down approach'' - where the neutrino oscillation parameters are predicted from a given  neutrino mass matrix and ``bottom-up approach''- where the existing neutrino oscillation data determine possible texture of the neutrino mass matrix. The first method``top-down approach'' relies on theoretical models associated with some discrete symmetries like $S_3, S_4, A_4, A_5$, etc.   \cite{puyam2022deviation,morisi2013neutrino,gonzalez2013s3} and the second method ``bottom up approach'' relies on the numerical analysis. We have used ``bottom-up approach'' to construct possible neutrino mass matrix numerically with the available oscillation data within 3$\sigma$ bound \cite{gonzalez2021nufit}. The presently available neutrino oscillation data includes two mass-squared difference i.e. solar mass-squared difference ($\Delta m_{21}^2$) and atmospheric mass-squared difference ($\Delta m_{32}^2$); three neutrino mixing angles i.e. solar mixing angle ($\theta_{12}$), reactor mixing angle ($\theta_{13}$) and atmospheric mixing angle ($\theta_{23}$) and one Dirac CP violating phase ($\delta_{CP}$) respectively. 

 A general complex Majorana neutrino mass matrix constructed with random input parameters, is diagonalized to generate an eigen-system which in turn predicts neutrino oscillation parameters consistent with the presently available oscillation data. The latest cosmological PLANCK's upper bound on the sum of three absolute neutrino mass eigenvalues, $\sum\vert m_i\vert<0.12$ eV        \cite{aghanim2020planck,tanseri2022updated} is used as a constraint in the numerical analysis. Some texture models\cite{nishiura1999lepton,grimus2004symmetry} are also possible to study numerically. One-zero texture and two-zero texture are studied in both normal hierarchy and inverted hierarchy within this latest cosmological bound.

\section{Adaptive Monte Carlo method for generating Majorana mass matrix}
The Dirac CP-violating phase can not be extracted from the real symmetric mass matrix. In order to extract the Dirac CP violating phase, the mass matrix should be a complex one i.e. all the elements of mass matrix are taken in the form of complex number as given below
\begin{equation}
m_{\nu}=\left(\begin{array}{ccc}
a_1+ia_2&b_1+ib_2&c_1+ic_2\\
b_1+ib_2&d_1+id_2&e_1+ie_2\\
c_1+ic_2&e_1+ie_2&f_1+if_2
\end{array}\right)
\label{mv}
\end{equation}

To construct numerically such a complex symmetric matrix, twelve real numbers are required i.e. a set of six real numbers for real part $\{a_1, b_1, c_1, d_1, e_1, f_1\}$ and another set of six real numbers for imaginary part $\{a_2, b_2, c_2, d_2, e_2, f_2\}$. These twelve real numbers are randomly generated within a given range, which are allowed by the latest cosmological PLANCK's upper bound, $\sum \vert m_i\vert<0.12$ eV. For a particular Majorana neutrino mass matrix, the extracted two mass-squared differences ($\Delta m_{21}^2 $ and $\Delta m_{32}^2$) and all the elements of neutrino mixing matrix, $U_{PMNS}$, are compared with the observed experimental values given in Table \ref{t1} \cite{esteban2019global,xing2021minimal,dziewit2006majorana}. The general structure of neutrino mixing matrix ($U_{PMNS}$) is given below:

\begin{equation}
U=U_{PMNS}=\left(\begin{array}{ccc}
U_{e1}&U_{e2}&U_{e3}\\
U_{\mu 1}&U_{\mu 2}&U_{\mu 3}\\
U_{\tau 1}&U_{\tau 2}&U_{\tau 3}
\end{array}\right)
\label{eq4}
\end{equation}
where, $\tan\theta_{12}=\frac{|U_{e2}|}{|U_{e1|}}$, $\tan\theta_{23}=\frac{|U_{\mu 3}|}{|U_{\tau 3}|}$, $\sin\theta_{13}=|U_{e3}|$.

The magnitude of the elements of  ($U_{PMNS}$) at $3\sigma$ ( i.e $99.77\%$ CL) is  given by the latest NuFIT 5.3(2024) \cite{gonzalez2021nufit}

 \begin{equation}
 |U|_{3\sigma}^{\mbox{w/o\ SK-atm}}= \left(\begin{array}{ccc}
 0.801\rightarrow 0.842&0.518\rightarrow 0.580&0.142\rightarrow 0.155\\
 0.236\rightarrow 0.507&0.458\rightarrow 0.691&0.630\rightarrow 0.779\\
 0.264\rightarrow 0.527&0.471\rightarrow 0.700&0.610\rightarrow 0.762
 \end{array}\right)
\end{equation}

The rejection rule for a considered mass texture to be a possible solution of eq (\ref{mv}) is defined as

\begin{equation}
\frac{(x_{i}^{cent}-x_{i})^{2}}{\sigma_{i}^{2}}\leq \frac{1}{\alpha^{2}}
\label{rr}
\end{equation}
where, $x_i^{cent}$ and  $x_i$ represent the experimentally observed central values and the numerically observed values of neutrino oscillation parameters respectively in our numerical analysis, while $\sigma_i$ and $\alpha$ are respectively the 3$\sigma$ uncertainties and a parameter which is used for fixing the assumed error $\sigma_i$.

 In the present analysis, we put $\alpha=1$ so that the results are obtained within 3$\sigma$ confidence level. If the extracted values of oscillation parameters from the mass matrix, are found to lie within the range of 3$\sigma$ bound, then it is one of the possible solutions to the general mass matrix of eq(\ref{mv}). To perform the numerical fitting of the mass matrix, we are using the Adaptive Monte Carlo method \cite{dziewit2006majorana,kiers2002ubiquitous}. This method also enables us to analyse a lot of possible solutions to the Majorana mass matrix including some zero-texture predicted by theoretical models.

The algorithm of the Adaptive Monte Carlo method is briefly described. It mainly works in two steps: (A) scattering  and (B) Adaptive Monte Carlo. In the scattering step-A: (i) it first randomly generates input parameters i.e. $a, b, c, d, e, f$ such that $a=a_1+ia_2, b=b_1+ib_2,...$; (ii) diagonalize the Hermitian matrix $h$ constructed by the $m_\nu$ to extract the neutrino oscillation parameters and (ii) compare the observed oscillation parameters with experimental data with the help of the $\chi_i^2$ function

$$
\chi_i^2=\frac{(x_{i}^{cent}-x_{i})^{2}}{(\frac{\sigma_{i}}{\alpha})^2}
$$.

\begin{table}
\renewcommand{\arraystretch}{1.5}
\centering
\begin{tabular}{ccccc} \hline
i&$x_{i}$&$x_{i}^{cent}$&$\sigma_{i}$ \\ 
1&$\Delta m_{31}^{2}[eV^2]$&$2.512\times10^{-3}$&$0.084\times 10^{-3}$ \\
2&$\Delta m_{21}^{2}[eV^2]$&$7.42\times10^{-5}$&$0.6\times 10^{-5}$ \\ 
3&$\vert U_{e1}\vert$&0.821&0.0205 \\ 
4&$\vert U_{e2}\vert$&0.549&0.031\\ 
5&$\vert U_{e3}\vert$&0.148&0.0065 \\ 
6&$\vert U_{\mu1}\vert$&0.371&0.135 \\ 
7&$\vert U_{\mu2}\vert$&0.574&0.116 \\ 
8&$\vert U_{\mu3}\vert$&0.704&0.074 \\ 
9&$\vert U_{\tau1}\vert$&0.395&0.131 \\ 
10&$\vert U_{\tau2}\vert$&0.585&0.114\\ 
11&$\vert U_{\tau3}\vert$&0.686&0.076 \\ \hline
\end{tabular}
\caption{ The best-fit or central absolute values of the neutrino mass-squared differences and the elements of $U_{PMNS}$ with the 3$\sigma$ uncertainties.} 
\label{t1}
\end{table}

The input parameters are allowed and saved if $\chi_{i}^2$ is less than the demanding number which is generally greater than one. In step-B: (i) Adaptive Monte Carlo first reads all possible points around each allowed points obtained in the scattering. The algorithm sets the range of input parameter for $n^{th}$ iteration (n=1,2,3,...,N) in such a way that $x_i^{cent}\pm\xi\delta$, 
where $\delta$ is an initial range and $\xi$ is a function defined as follows

$$
\xi=\left\{\begin{array}{lll}
1, &\text{if}& n=0\\
0.6/n, & \text{if} &n>0
\end{array}\right\}.
$$

The program follows the three primary tasks i.e random number generation, diagonalization and comparison with experimental data for N number of iterations; (ii) it finally chooses the set of input parameters for which the calculated value of $\chi^2$ is the lowest,

$$
\chi^2=\sum^{11}_{i=1}\chi_i^2.
$$

This set of input parameters is taken as ``the best set'' and it gives new central value $x_i^{cent}$ as observed output data.

\begin{table}
\renewcommand{\arraystretch}{1.5}
\centering
 \begin{tabular}{ccccc}
\hline
Parameter&I&II&III&IV\\ \hline
$\Delta m_{31}^{2}[eV^2]$&$2.52\times10^{-3}$&$2.48\times10^{-3}$&$2.44\times10^{-3}$&$2.49\times10^{-3}$ \\  
$\Delta m_{21}^{2}[eV^2]$&$7.38\times10^{-5}$&$7.34\times10^{-5}$&$7.52\times10^{-5}$&$7.32\times10^{-5}$ \\ 
$\vert m_{1}\vert$&0.02350&0.01747&0.01983&0.00741 \\ 
$\vert m_{2}\vert$&0.02503&0.01946&0.02165&0.01131\\ 
$\vert m_{3}\vert$&0.05543&0.05280&0.05325&0.05045 \\ 
$\vert U_{e1}\vert$&0.8223&0.8286&0.8175&0.8255 \\ 
$\vert U_{e2}\vert$&0.5491&0.5397&0.5560&0.5441\\ 
$\vert U_{e3}\vert$&0.1492&0.1483&0.1493&0.1493 \\ 
$\vert U_{\mu1}\vert$&0.4284&0.4198&0.4203&0.4660\\ 
$\vert U_{\mu2}\vert$&0.5350&0.5942&0.6001&0.5958 \\ 
$\vert U_{\mu3}\vert$&0.7281&0.6859&0.6805&0.6540 \\ 
$\vert U_{\tau1}\vert$&0.3745&0.3702&0.3935&0.3181 \\ 
$\vert U_{\tau2}\vert$&0.6420&0.5962&0.5749&0.5906 \\ 
$\vert U_{\tau3}\vert$&0.6689&0.7123&0.7173&0.7415 \\ 
$\theta_{12}[/^0]$&33.73&33.07&34.22&33.39 \\ 
$\theta_{13}[/^0]$&8.58&8.53&8.58&8.59 \\ 
$\theta_{23}[/^0]$&47.42&43.92&43.49&41.40\\ 
$\delta_{_{CP}}[/^0]$&240.32&257.87&267.52&233.98 \\ 
$\alpha_1[/^0]$&175.60&31.1&23.47&86.57\\ 
$\alpha_2[/^0]$&585.72&317.91&103.52&107.7 \\ 
$m_{_{\beta\beta}} [eV]$&0.04513&0.04177&0.4221&0.03683 \\
$m_{_\beta} [eV]$&0.04752&0.04485&0.04503&0.04188\\
$\sum |m_i|[eV]$&0.1039&0.08974&0.09474&0.06918  \\            \hline
\end{tabular}
\caption{The best central values of the neutrino oscillation parameters within 3$\sigma$ bound of experimental data calculated by using Adaptive Monte Carlo method in NH. The roman figures I, II, III and IV represent the numerical results obtained in four different set of runs. All the neutrino oscillation parameters are found within 3$\sigma$ bound of experimental data consistent with the latest cosmological PLANCK's upper mass bound $\sum|m_i|<0.12$ eV.} 
\label{nhamc}
\end{table}

\begin{table}
\renewcommand{\arraystretch}{1.5}
\centering
 \begin{tabular}{ccccc}
\hline
Parameter&$a=0$&$b=0$&$c=0$\\ \hline
$\Delta m_{31}^{2}[eV^2]$&$2.46\times10^{-3}$&$2.48\times10^{-3}$&$2.49\times10^{-3}$ \\  
$\Delta m_{21}^{2}[eV^2]$&$7.33\times10^{-5}$&$7.29\times10^{-5}$&$7.41\times10^{-5}$ \\ 
$\vert m_{1}\vert$&0.00722&0.01138&0.00729  \\ 
$\vert m_{2}\vert$&0.01120&0.01423&0.01128  \\ 
$\vert m_{3}\vert$&0.05020&0.05110&0.05044   \\ 
$\vert U_{e1}\vert$&0.8079&0.8241&0.8181    \\ 
$\vert U_{e2}\vert$&0.5709&0.5464&0.5555    \\ 
$\vert U_{e3}\vert$&0.1459&0.1490&0.1484    \\ 
$\vert U_{\mu1}\vert$&0.4119&0.4176&0.3020  \\ 
$\vert U_{\mu2}\vert$&0.5745&0.6087&0.6224   \\ 
$\vert U_{\mu3}\vert$&0.7072&0.6745&0.7220   \\ 
$\vert U_{\tau1}\vert$&0.4213&0.3826&0.4893  \\ 
$\vert U_{\tau2}\vert$&0.5864&0.5751&0.5513   \\ 
$\vert U_{\tau3}\vert$&0.6917&0.7230&0.6757   \\ 
$\theta_{12}[/^0]$&35.24&33.54&34.17             \\ 
$\theta_{13}[/^0]$&8.38&8.56&8.53          \\ 
$\theta_{23}[/^0]$&45.63&43.00&46.89           \\ 
$\delta_{_{CP}}[/^0]$&269.67&266.09&201.24        \\ 
$\alpha_1[/^0]$&68.40&93.57&48.44         \\ 
$\alpha_2[/^0]$&695.31&192.58&46.89        \\ 
$m_{_{\beta\beta}}[eV]$&0.03536&0.03842&0.03626 \\
$m_{_\beta}[eV]$&0.04080&0.04262&0.04150 \\
$\sum |m_i|[eV]$&0.06862&0.07672&0.06903 \\ \hline
\end{tabular}
\caption{The best central values of the neutrino oscillation parameters for one zero texture allowed in NH within 3$\sigma$ bound of experimental data calculated by using Adaptive Monte Carlo method. All the neutrino oscillation parameters are found within 3$\sigma$ bound of experimental data consistent with the latest cosmological PLANCK's upper mass bound $\sum|m_i|<0.12$ eV.} 
\label{zt1namc}
\end{table}

\section{Numerical analysis and results}

We find several possible solutions for the general Majorana neutrino mass matrix by Adaptive Monte Carlo method. In the first step, possible rough solutions for the mass matrix, are searched by giving a suitable demanding number on the right side of rejection rule in eq (\ref{rr}). A certain number of possible solutions are taken for consideration of numerical analysis and each of possible solutions is represented by a set of points. If the demanding number is equal to one, then all possible solutions are found within 3$\sigma$ bound of experimental data. In the second step, the algorithm finds a large numbers of possible solutions about the surrounding of each of the points in the first step, and then select the best one by using Adaptive Monte Carlo (AMC) method.

In the numerical analysis of general Majorana mass matrix, we first find the solutions for all non-zero elements. The two neutrino hierarchical mass ordering i.e, normal hierarchy ($m_3>m_2>m_1$) and inverted hierarchy ($m_2>m_1>m_3$) are considered in the analysis. The sum of the three absolute mass eigenvalues given by latest cosmological PLANCK's data, $\sum\vert m_i\vert<0.12$ eV is used as a constraint in the generation of neutrino mass matrix. This cosmological upper mass bound allows only normal hierarchy (NH) consistent with experimentally observed neutrino oscillation data in case of neutrino mass structure whose elements are all non-zero. The numerical values of oscillation parameters obtained in NH are presented in Table \ref{nhamc}. The numerical results are given by the four set of runs in four columns represented by four roman figures I, II, III and IV. The numerical results in our analysis are confirmed by the random number generation of $10^8$ iterations in the range from -0.04 to +0.04 for each elements of the mass matrix. The wide range of the elements farther away from a preferable range requires large number of iterations and it always converges to the allowed range. The range of the elements has a limit towards the smaller values which is constrained by the sum of three absolute neutrino masses.

\begin{table}
\renewcommand{\arraystretch}{1.5}
\centering
 \begin{tabular}{ccccc}
\hline
Parameter&a, b=0&a, c=0\\ \hline
$\Delta m_{31}^{2}[eV^2]$&$2.50\times10^{-3}$&$2.50\times10^{-3}$ \\ 
$\Delta m_{21}^{2}[eV^2]$&$7.54\times10^{-5}$&$7.44\times10^{-5}$ \\ 
$\vert m_{1}\vert$&0.005738&0.0055  \\ 
$\vert m_{2}\vert$&0.01036&0.01023  \\ 
$\vert m_{3}\vert$&0.05041&0.05033   \\ 
$\vert U_{e1}\vert$&0.8184&0.8188    \\ 
$\vert U_{e2}\vert$&0.5553&0.5546    \\ 
$\vert U_{e3}\vert$&0.1475&0.1476    \\ 
$\vert U_{\mu1}\vert$&0.3615&0.4343  \\ 
$\vert U_{\mu2}\vert$&0.6258&0.5468  \\ 
$\vert U_{\mu3}\vert$&0.6911&0.7157   \\ 
$\vert U_{\tau1}\vert$&0.4466&0.3751  \\ 
$\vert U_{\tau2}\vert$&0.5476&0.6271   \\ 
$\vert U_{\tau3}\vert$&0.7075&0.6826   \\ 
$\theta_{12}[/^0]$&34.15&34.11             \\ 
$\theta_{13}[/^0]$&8.48&8.83           \\ 
$\theta_{23}[/^0]$&44.32&46.35            \\ 
$\delta_{_{CP}}[/^0]$&236.41&243.01        \\ 
$\alpha_1[/^0]$&34.15&150.74        \\ 
$\alpha_2[/^0]$&44.32&443.05        \\
$m_{_{\beta\beta}} [eV]$&0.03576&0.03567 \\
$m_{_\beta} [eV]$&0.041412&0.04136\\ 
$\sum |m_i|[eV]$&0.06652&0.06608 \\ \hline
\end{tabular}
\caption{The best central values of the neutrino oscillation parameters for two zero texture allowed in NH within 3$\sigma$ bound of experimental data calculated by using Adaptive Monte Carlo method. All the neutrino oscillation parameters are found within 3$\sigma$ bound of experimental data consistent with the latest cosmological PLANCK's upper mass bound $\sum|m_i|<0.12$ eV.} 
\label{zt2namc}
\end{table}
 
All the elements in the general Majorana mass matrix can not be exactly predicted by the theoretical model. Our numerical analysis can give a rough ranges of elements of the mass matrix which are constrained by the low energy neutrino oscillation data within 3$\sigma$ bound. The absolute values of the elements in the general complex symmetric mass matrix give value of elements of Majorana mass matrix. The elements of Majorana mass matrix are given by $m_{11}=\vert a \vert$, $m_{12}=\vert b \vert$, $m_{13}=\vert c \vert$, $m_{22}=\vert d \vert$, $m_{23}=\vert e \vert$, and $m_{33}=\vert f \vert$ respectively. The numerical values in normal hierarchical mass models is constrained by the latest cosmological PLANCK's upper bound, $\sum \vert m_i \vert<0.12$ eV.

The allowed range and best fit values of all mass elements of the Majorana mass matrix and observable mass bounds calculated by the Adaptive Monte Carlo method are given in Table \ref{mrange}. From the analysis of the numerical values given in Table \ref{mrange}, we observe that the numerical values of all mass elements are found at the order of $10^{-3}-10^{-2}$. In the NH mass model, the best fit numerical values of the mass elements $m_{11}$, $m_{12}$, and $m_{13}$ are generally found near the order of $10^{-3}$ whereas the best fit numerical values of the mass elements $m_{22}$, $m_{23}$ and $m_{33}$ are also found near the order of $10^{-2}$. We can also calculate the other two important parameters in neutrino physics namely the effective neutrino mass $m_{\beta \beta}$ of neutrinoless double beta decay ($0\nu\beta\beta$) and effective electron neutrino mass $m_{\beta}$ from tritium beta decay using the following relations:

\begin{equation}
m_{_{\beta\beta}}=\vert \sum^3_{i=1} U_{ei}^2 m_i\vert,\hspace{0.5cm} i=1,2,3;
\end{equation}
and
\begin{equation}
m_{_\beta}=\left[\sum^3_{i=1} \vert U_{ei} \vert^2 m_i^2\right]^{\frac{1}{2}};\hspace{0.5cm} i=1,2,3.
\end{equation}
The latest experimental upper bound on effective neutrino mass  and effective electron mass are given by $m_{\beta\beta}<(0.036-0.156)$ eV  \cite{abe2024search,agostini2013results,agostini2020final,agostini2018improved,ackermann2013gerda} and $m_{\beta}<0.8$ eV   \cite{aker2019improved,aker2020first,aker2021design,aker2022katrin} respectively.

This numerical analysis also enables us to analyse some zero textures predicted by theoretical model. We study the special case for one-zero texture i.e. $a=0$, $b=0$, $c=0$, $d=0$, $e=0$ and $f=0$ in both NH and IH. Only three cases of one-zero texture i.e. $a=0$, $b=0$ and $ c=0$, are allowed in NH whereas none of one-zero texture structure is allowed. The observed neutrino oscillation parameters obtained in one-zero texture for NH are presented in Table \ref{zt2namc}.  Two-zero texture are mainly studied numerically. For $3\times3$ matrix, there are fifteen possible cases for two-zero texture. Out of fifteen cases only two case of texture for $a,b=0$ and $a,c=0$, are allowed in normal hierarchy. None of the two-zero textures is also found as a solution for neutrino mass matrix. The one-zero texture and two-zero texture allowed by NH have the following mass structures:

\begin{align}
m_{\nu} &=\left(\begin{array}{ccc}
0&b&c\\
b&d&e\\
c&e&f
\end{array}\right),\;
m_{\nu}=\left(\begin{array}{ccc}
a&0&c\\
0&d&e\\
c&e&f
\end{array}\right),\;
m_{\nu}=\left(\begin{array}{ccc}
a&b&0\\
b&d&e\\
0&e&f
\end{array}\right), \nonumber \\ 
m_{\nu} &=\left(\begin{array}{ccc}
0&0&c\\
0&d&e\\
c&e&f
\end{array}\right), \;
m_{\nu}=\left(\begin{array}{ccc}
0&b&0\\
b&d&e\\
0&e&f
\end{array}\right).
\end{align}

The observed neutrino oscillation parameters obtained in NH for the cases $ a,b=0$ and $a,c=0$ are given in Table \ref{zt2namc}.

\begin{figure}
\subfigure[]{
\includegraphics[width=.46\textwidth]{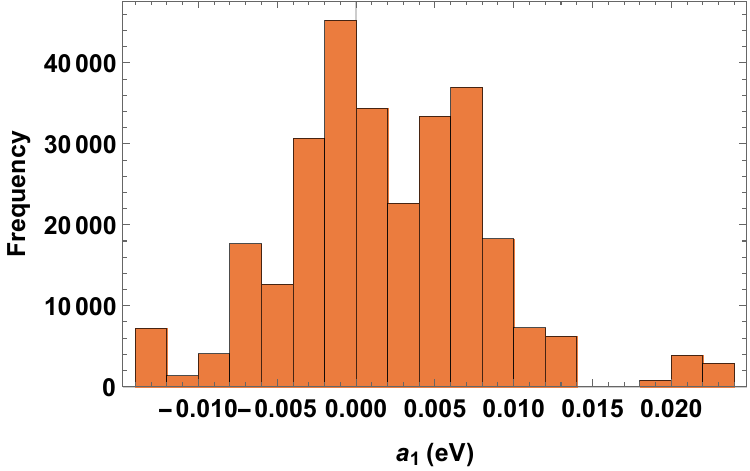}}
\quad
\subfigure[]{
\includegraphics[width=.46\textwidth]{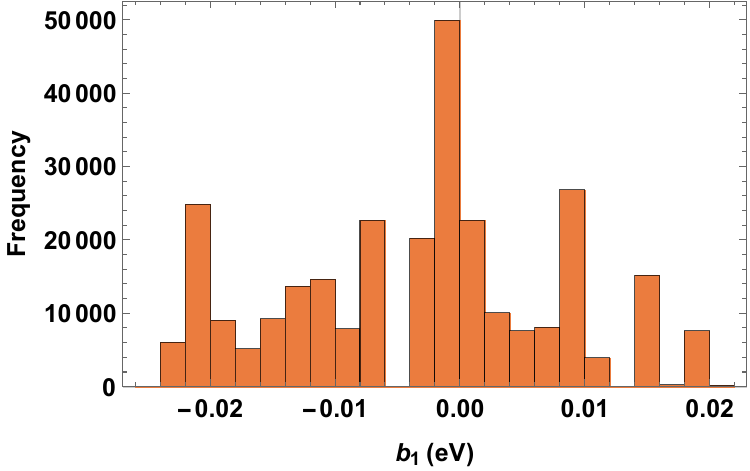}}
\quad
\subfigure[]{
\includegraphics[width=.46\textwidth]{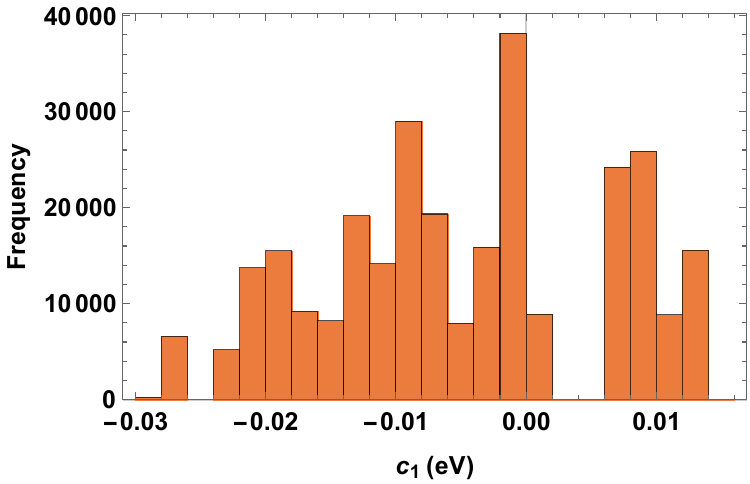}}
\quad
\subfigure[]{
\includegraphics[width=.46\textwidth]{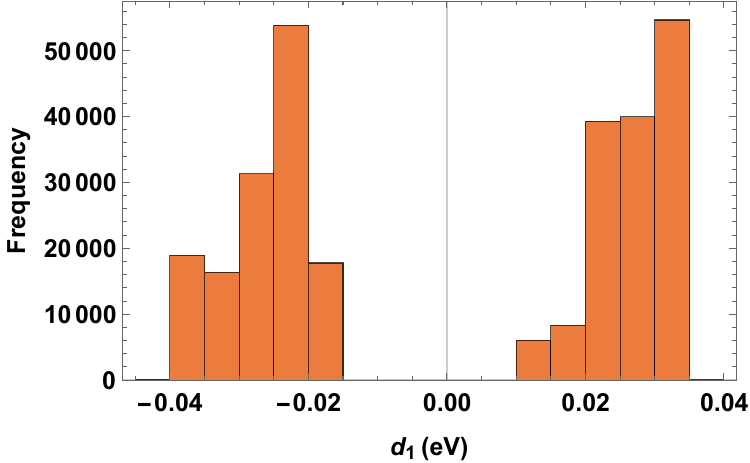}}
\quad
\subfigure[]{
\includegraphics[width=.46\textwidth]{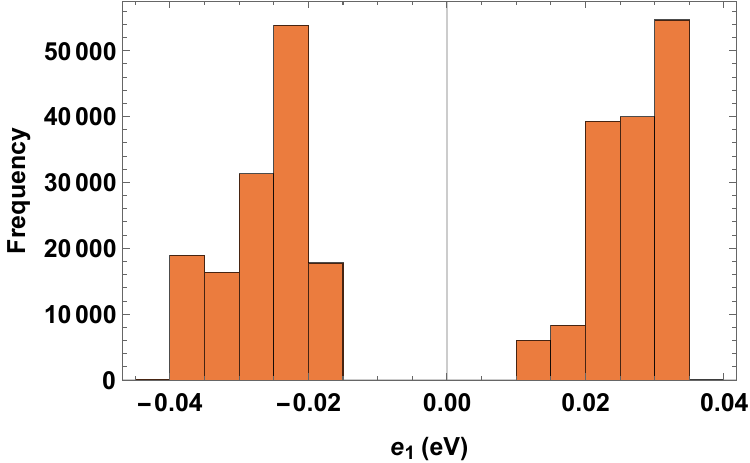}}
\quad
\subfigure[]{
\includegraphics[width=.46\textwidth]{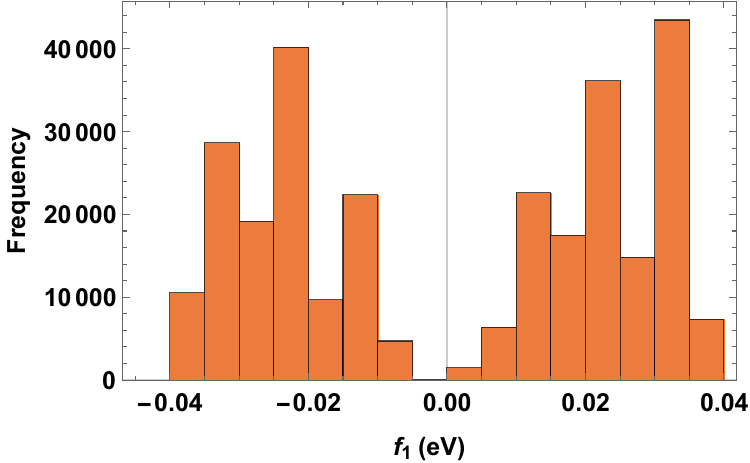}}
\quad
\caption{Frequency spectrum for the elements of the neutrino mass matrix in normal mass hierarchy (NH). The numerical values in region of higher peak of frequency spectrum shows more possible solutions as compared to lower peak region of the frequency spectrum.}
\label{fsnh}
\end{figure}

When the set of six real numbers for imaginary part $\{a_2, b_2, c_2, d_2, e_2, f_2\}$ is equal to zero, then complex symmetric matrix becomes a real symmetric matrix. For such real symmetric Majorana mass matrix, we also estimate all the elements of the mass matrix for normal hierarchy (NH) and we then plot their frequency spectrum.  These frequency spectrum are the plot of number of allowed points versus the allowed range of the elements of the mass matrix.  That is, the higher peak represents the lager number of points whereas the lower peak represents the smaller number of points. The physical interpretation of frequency spectrum for the elements of the mass matrix is given for normal hierarchy (NH) in Figure \ref{fsnh}.

\begin{figure}
\subfigure[]{
\includegraphics[width=.46\textwidth]{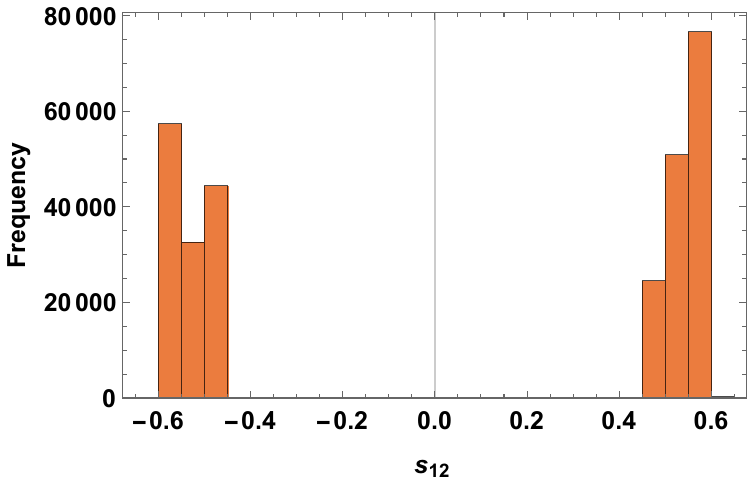}}
\quad
\subfigure[]{
\includegraphics[width=.46\textwidth]{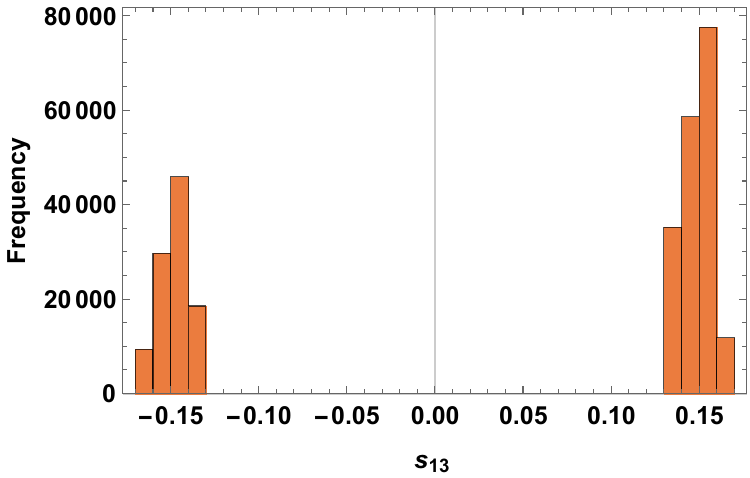}}
\quad
\subfigure[]{
\includegraphics[width=.46\textwidth]{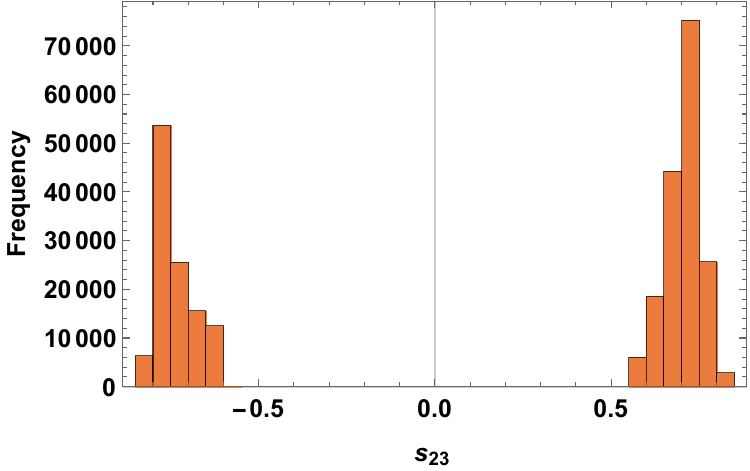}}
\quad
\subfigure[]{
\includegraphics[width=.46\textwidth]{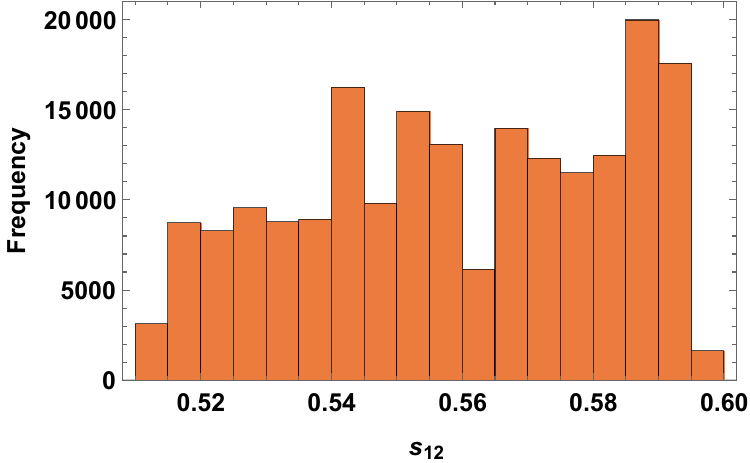}}
\quad
\subfigure[]{
\includegraphics[width=.46\textwidth]{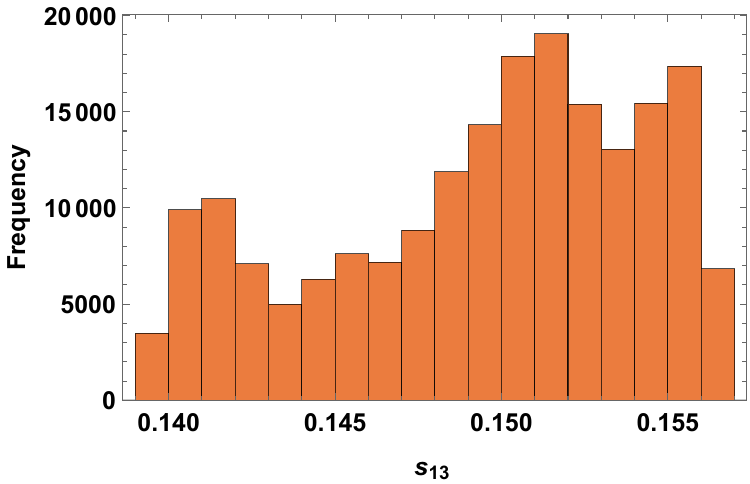}}
\quad
\subfigure[]{
\includegraphics[width=.46\textwidth]{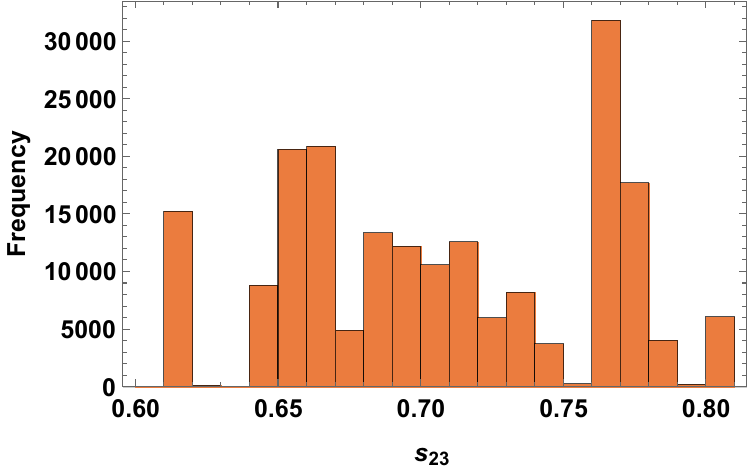}}
\quad
\caption{ \footnotesize (a)-(c) represent frequency spectrum of sine of the three mixing angles allowed by the mixing matrix in normal hierarchy (NH).  (d)-(f) represent frequency spectrum of absolute values of the sine of the three mixing angles obtained from the mixing matrix.}
\label{fsma}
\end{figure}

From \ref{fsnh}, we observe that the frequency spectrum for the elements $a=a_1$, $b=b_1$, and $c=c_1$ shows preferable very small values nearer to centre from both sides on x-axis. The frequency spectrum for the elements $d=d_1$, $e=e_1$, and $f=f_1$ shows preferable small non-zero values farther away from the centre of both sides on x-axis i.e. both positive and negative values are possible. The sine of three mixing angles found in NH shows two possible allowed frequency peaks on both sides of x-axis, which are shown in Figure \ref{fsma}. The negative values of the sine of mixing angles are due to presence of negative sign in the elements of the observed mixing matrix ($U_{PMNS}$) and its positive mixing angles can be obtained by taking absolute values of all the elements of the observed mixing matrix. The frequency spectrum of sine of mixing angles obtained from the mixing matrix and absolute values mixing angles are shown in Figure \ref{fsma}.

\begin{table}
\renewcommand{\arraystretch}{1.5}
\centering
 \begin{tabular}{cccc}
\hline 
Parameters (NH)&Range&Best fit value\\ \hline
$m_{11}$&0.001136-0.02736&0.01111 \\ 
$m_{12}$&$5.33\times10^{-5}$-0.02239&0.0074 \\ 
$m_{13}$&0.003061-0.01840&0.01593 \\ 
$m_{22}$&0.01696-0.03040&0.02407 \\ 
$m_{23}$&0.02264-0.04120&0.03201 \\ 
$m_{33}$&0.00388-0.02835&0.01689 \\
$ m_{_{\beta \beta}}$\;[eV]&0.03368-0.04943&0.04235  \\ 
$ m_{_\beta}$\;[eV]&0.04011-0.05122&0.04534  \\ 
$\sum \vert m_i \vert$;[eV]&0.06128-0.1163&0.09325   \\ \hline
\end{tabular} 
\caption{ The absolute values of allowed range and the best fit value of the neutrino mass elements in normal hierarchical mass model consistent with experimental data within 3$\sigma$ uncertainties and the latest PLANCK's mass bound $\sum|m_i|<0.12$ eV.}
\label{mrange}
\end{table}

\begin{figure}
\subfigure[]{
\includegraphics[width=.46\textwidth]{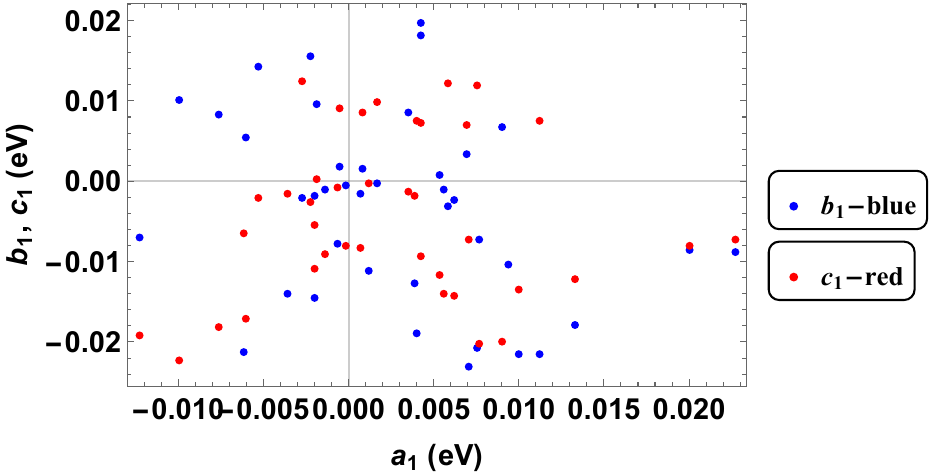}}
\quad
\subfigure[]{
\includegraphics[width=.46\textwidth]{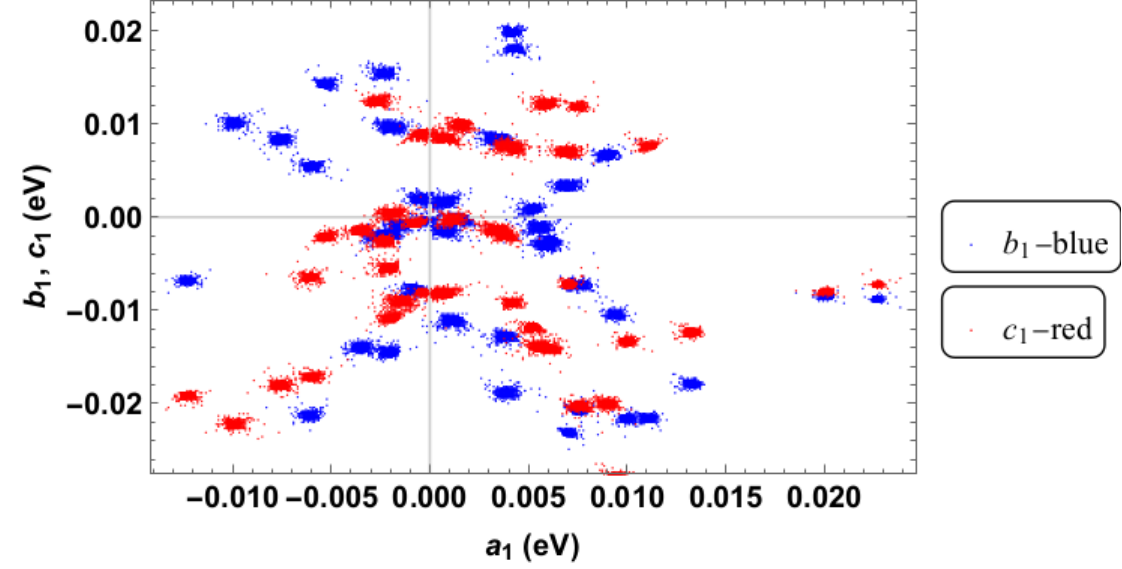}}
\quad
\subfigure[]{
\includegraphics[width=.46\textwidth]{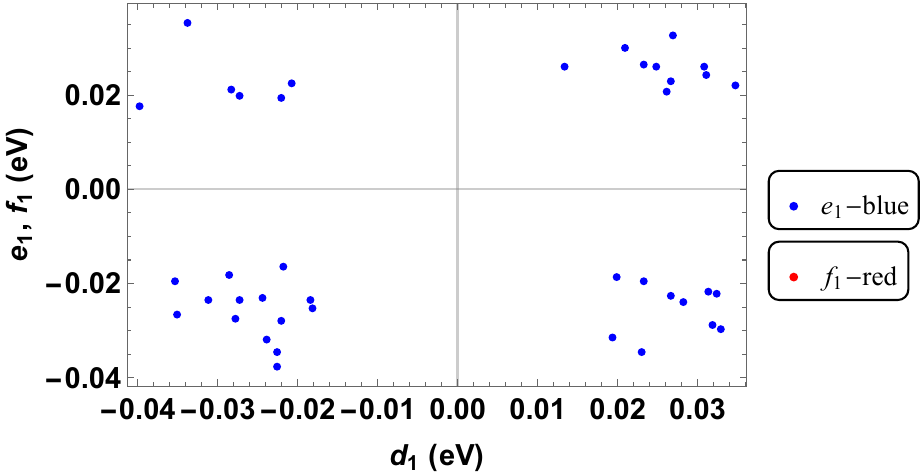}}
\quad
\subfigure[]{
\includegraphics[width=.46\textwidth]{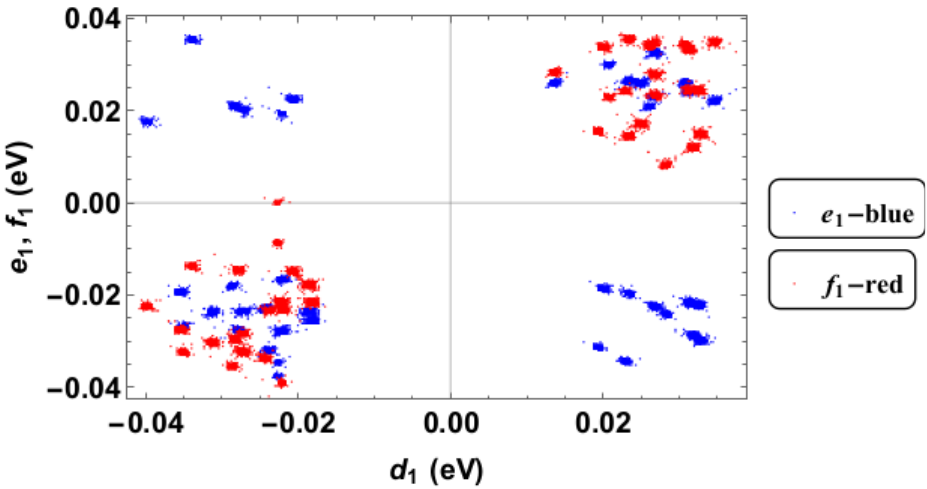}}
\quad
\caption{ \footnotesize The graphical plots of allowed input elements $a, b, c, d, e$ and $f$ of real Majorana mass texture generated by using Adaptive Monte Carlo simulation. Thin and thick plots of the above Figure represent scatter plots for rough solutions and enlarged regions for best the fit values in AMC method for the Majorana neutrino mass matrix in normal hierarchy(NH).}
\label{scadplots}
\end{figure}

The graphical scatter plots of the set of elements $\{a_1, b_1, c_1\}$ and $\{d_1, e_1, f_1\}$ for rough solutions and  the enlarged regions obtained in the AMC mthod for the above corresponding elements in NH are shown in the Figure \ref{scadplots}. These plots are drawn by using the points obtained for each elements in a single run. For such real symmetric Majorana mass matrix, we also study two zero texture classified as 

\begin{align*}
A_1:&\ a,\ b=0; \\ 
A_2:&\ a,\ c=0; \\
B_1:&\ c,\ d=0; \\
B_2:&\  b,\ f=0; \\
B_3:&\ b,\ d=0; \\
B_4:&\ c,\ f=0 ;\ \mbox{and} \\ 
C:&\ d,\ f=0. 
\end{align*}

These two-zero texture are allowed with the earlier cosmological PLANCK's upper bound, $\sum \vert m_i \vert<0.23$ eV \cite{collaboration2015planck,ade2016planck} in both NH and IH mass models.

\section{Summary and conclusion}

To summarize, we use the Adaptive Monte Carlo method for numerical analysis where the elements of Majorana neutrino mass matrix are randomly generated. The Majorana neutrino mass matrix should be complex matrix in order to account for the CP violating Dirac phase. A complex symmetric mass matrix is essentially considered for CP non-conserving case. Twelve parameters are required to construct a complex symmetric Majorana mass matrix. These twelve parameters correspond to three mass eigenvalues ($m_1, m_2, m_3$), three mixing angles ($\theta_{12}, \theta_{13}, \theta_{23}$), three CP violating phases ($\delta_{_{CP}}, \alpha_1, \alpha_2$) and three non-physical phases ($\phi_1, \phi_2, \phi_3$) which can be absorbed by the charged fermion Dirac fields.

In numerical analysis, twelve parameters are represented by real numbers in such a way that $a=a_1+ia_2, b=b_1+ib_2, c=c_1+ic_2, d=d_1+id_2, e=e_1+ie_2, f=f_1+if_2$. The set of twelve real numbers is randomly generated to construct a general Majorana mass matrix structure. A Hermitian matrix is formed by using the complex symmetric mass matrix so that direct digonalization of the Hermitian matrix leads to the exact mixing matrix constructed by the eigenvectors of the Hermitian matrix and the corresponding mass-squared eigenvalues. The main result of the numerical analysis shows only normal hierarchical mass models with all non-zero elements are valid within 3$\sigma$ bound of experimental data, consistent with the latest PLANCK's upper mass bound, $\sum \vert m_i\vert<0.12$ eV. Some studies \cite{di2021most,di2022neutrino} have also indicated the most constraining bound to date, $\sum\vert m_i\vert<0.09$ eV.

The present numerical analysis is also extended to study some zero texture. The one-zero texture for $a=0$, $b=0$ and $c=0$ are only allowed in NH mass model whereas none of one-zero texture is allowed in IH mass model. Two zero texture are also studied in both normal and inverted hierarchical mass models. A $3\times3$ Majorana neutrino mass structure can have fifteen two-zero texture. Out of the fifteen possible zero texture, only two cases for $a,b=0$ and $a,c=0$ are allowed in normal hierarchical mass model. The same viability of these allowed two-zero texture is also reported \cite{denton2024survey}. None of these two zero texture is allowed in inverted hierarchical mass model. The present finding may have important implications for model building using discrete flavour symmetries \cite{king2013neutrino,smirnov2013neutrino,altarelli2010discrete}.

\bibliographystyle{unsrt}
\bibliography{mmm}
\end{document}